\documentclass[twoside,twocolumn,9pt]{article}
\usepackage{extsizes}
\usepackage[super,sort&compress,comma]{natbib} 
\usepackage[version=3]{mhchem}
\usepackage[left=1.5cm, right=1.5cm, top=1.785cm, bottom=2.0cm]{geometry}
\usepackage{balance}
\usepackage{times,mathptmx}
\usepackage{sectsty}
\usepackage{graphicx} 
\usepackage{lastpage}
\usepackage[format=plain,justification=justified,singlelinecheck=false,font={stretch=1.125,small,sf},labelfont=bf,labelsep=space]{caption}
\usepackage{float}
\usepackage{fancyhdr}
\usepackage{fnpos}
\usepackage[english]{babel}
\usepackage{array}
\usepackage{droidsans}
\usepackage{charter}
\usepackage[T1]{fontenc}
\usepackage[usenames,dvipsnames]{xcolor}
\usepackage{setspace}
\usepackage[compact]{titlesec}

\usepackage{subcaption}
\usepackage{SIunits}

\definecolor{linkcolor}{rgb}{0,0,0.6} 
\usepackage[ pdftex,colorlinks=true,
pdfstartview=FitV,
linkcolor= linkcolor,
citecolor= linkcolor,
urlcolor= linkcolor,
hyperindex=true,
hyperfigures=false]
{hyperref} 

\usepackage{fancyhdr} 
\usepackage{enumitem}

\usepackage{xr}
\usepackage{epstopdf}

\definecolor{cream}{RGB}{222,217,201}
\begin{document}

\pagestyle{fancy}
\thispagestyle{plain}


\makeFNbottom
\makeatletter
\renewcommand\LARGE{\@setfontsize\LARGE{15pt}{17}}
\renewcommand\Large{\@setfontsize\Large{12pt}{14}}
\renewcommand\large{\@setfontsize\large{10pt}{12}}
\renewcommand\footnotesize{\@setfontsize\footnotesize{7pt}{10}}
\makeatother

\renewcommand{\thefootnote}{\fnsymbol{footnote}}
\renewcommand\footnoterule{\vspace*{1pt}%
\color{cream}\hrule width 3.5in height 0.4pt \color{black}\vspace*{5pt}} 
\setcounter{secnumdepth}{5}

\makeatletter 
\renewcommand\@biblabel[1]{#1}            
\renewcommand\@makefntext[1]%
{\noindent\makebox[0pt][r]{\@thefnmark\,}#1}
\makeatother 
\renewcommand{\figurename}{\small{Fig.}~}
\sectionfont{\sffamily\Large}
\subsectionfont{\normalsize}
\subsubsectionfont{\bf}
\setstretch{1.125} 
\setlength{\skip\footins}{0.8cm}
\setlength{\footnotesep}{0.25cm}
\setlength{\jot}{10pt}
\titlespacing*{\section}{0pt}{4pt}{4pt}
\titlespacing*{\subsection}{0pt}{15pt}{1pt}


%
%
%

\makeatother

\twocolumn[
  \begin{@twocolumnfalse}
\sffamily
\begin{tabular}{m{1.5cm} p{13.5cm} }

\vspace{0.3cm} & \noindent\LARGE{\textbf{The surface tells it all: Relationship between volume and surface fraction of liquid dispersions$^\dag$}} \\
\vspace{0.3cm} & \vspace{0.3cm} \\

\vspace{0.3cm} & \noindent\large{Emilie Forel,$^{\ast}$\textit{$^{a}$} Emmanuelle Rio,\textit{$^{a}$} Maxime Schneider,\textit{$^{a}$} Sebastien Beguin,\textit{$^{a}$} Denis Weaire,\textit{$^{b}$} Stefan Hutzler,\textit{$^{b}$} and Wiebke Drenckhan\textit{$^{a}$}} \\

\vspace{0.3cm} & \noindent\normalsize{The properties of liquid dispersions, such as foams or emulsions, depend strongly on the volume fraction $\phi$ of the continuous phase. Concentrating on the example of foams, we show experimentally and theoretically that $\phi$ may be related to the fraction $\phi_s$ of the surface at a wall which is wetted by the continuous phase - given an expression for the interfacial energy or osmotic pressure of the bulk system. Since the surface fraction $\phi_s$ can be readly determined from optical measurement and since there are good general approximations available for interfacial energy and osmotic pressure we thus arrive at an advantageous method of estimating $\phi$. The same relationship between $\phi$ and $\phi_s$ is also expected to provide a good approximation of the fraction of the bubble or drop surface which is wetted by the continuous phase. This is a parameter of great importance for the rheology and ageing of liquid dispersions.} \\

\end{tabular}

 \end{@twocolumnfalse} \vspace{0.6cm}

  ]

\renewcommand*\rmdefault{bch}\normalfont\upshape
\rmfamily
\section*{}
\vspace{-1cm}


\footnotetext{\textit{$^{a}$~Address, Laboratoire de Physique des Solides, CNRS, Univ. Paris-Sud, Universit\'e Paris-Saclay, 91405 Orsay Cedex, France.  E-mail: emilie.forel@u-psud.fr}}
\footnotetext{\textit{$^{b}$~Address,School of Physics, Trinity College Dublin, The University of Dublin, Ireland. }}

\footnotetext{\dag~Electronic Supplementary Information (ESI)}



\section{Introduction}



Liquid dispersions, such as foams or emulsions, consist of individual gas bubbles or liquid droplets, respectively, which are dispersed in a continuous liquid matrix. The volume fraction $\phi$ of the continuous phase ranges from zero in the limit of tightly compacted, polyhedral bubbles/droplets, to a critical value $\phi_c$, in the limit where the bubbles/droplets are spherical and come apart \cite{Weaire2001,Cantat2013,VanHecke2010}. 
 Most physical properties depend strongly on $\phi$ - for example, the shear modulus of disordered liquid dispersions decreases to zero at $\phi_c$ \cite{Cantat2013,Hohler2005,Weaire2001,Princen2001,Bink1998}.

A number of different experimental techniques exist for the determination of the value of $\phi$, based, for example, on the measurement of  electrical conductivity \cite{Feitosa2005} or capacity \cite{Hutzler1995}, optical transmission \cite{Vera2001} or X-ray techniques \cite{Solorzano2013}. Here we show that $\phi$ is readily determined from a simple optical measurement of the surface fraction $\phi_s$, that is, the fractional area of the continuous liquid phase which wets the surface of the vessel holding the dispersion. We will show that the relationship between the volume and the surface fraction depends on the interfacial energy $E$ of the dispersion which is proportional to the interfacial area created between the immiscible fluids. Various expressions exist for $E$ in the literature (\cite{Drenckhan2015} and references therein) based on approximate theory or semi-empirical fits. 

Previous work on surface properties of foams has focussed on the relationship between the bubble size or the bubble size distribution of the surface and the bulk structure \cite{Cheng1983, Wang2009}. Here we concentrate entirely on the question of the relationship between the volume and surface fraction, $\phi$ and $\phi_s$, which is quite insensitive to the detailed structure of the dispersions, as will be shown later. In what follows we will refer to bubbles making up a foam. However, the nature of the physical argument and comparison with data on emulsions from the literature shows the general validity of our proposed relations.

\section{Experimental set-up}
 \begin{figure}[ht!]
		\includegraphics[width=8cm]{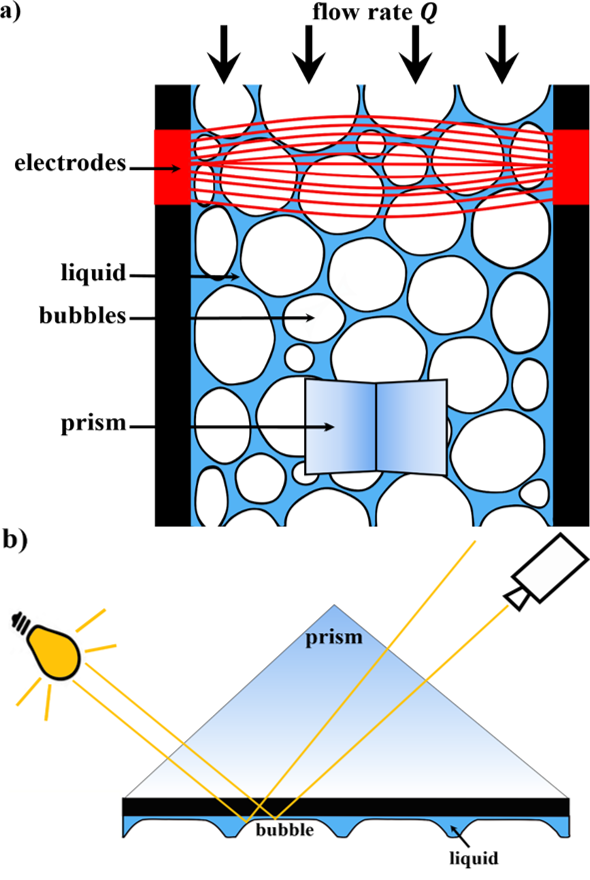} 
		\caption{(a) Scheme of the experimental setup: the foam is contained in a glass column of square cross-section (FoamScan by TECLIS). A pair of electrodes (in red) is used to measure the local electrical conductivity of the foam (and hence the volume fraction $\phi$).  Foaming solution is injected at constant, controlled flow rate $Q$ in order to ensure a homogeneous volume fraction throughout the foam. (b)Ray paths passing through a prism are used to observe bubbles at the wall.}
		\label{fig:schema}
\end{figure}
In the experiments we use foams generated from aqueous solutions of the anionic surfactant Sodium Dodecyl Sulfate (SDS, $6~\gram\per\liter$) with some additional glycerol ($0.5~\rm{wt}\%$) and dodecanol ($0.06~\gram\per\liter$) in order to improve foam and solution stability.
In order to obtain foams over a wide range of of different bubble radii  $R$ ($0.1 <\langle R\rangle< 2~\milli\meter$) and polydispersities $\sigma = \frac{\sqrt{\langle R^2 \rangle - \langle R\rangle ^2}}{\langle R\rangle}$ ($1~-64~\%$) we use different foaming techniques (sec.$~$\ref{app:generation} and \ref{app:size_poly} in the Supplementary Materials). \textit{Monodisperse} foams ($\sigma<~2\%$) are generated using microfluidic techniques or bubbling from a needle. \textit{Polydisperse} foams of low polydispersity are generated by bubbling from a porous sparger ($\sigma\approx 5~\%$), while larger polydispersities are obtained by the double syringe method \cite{Drenckhan2015} ($\sigma \approx 40~\%$ ) or by shaking the solutions in a closed container ($\sigma \approx 50~\%$ ).

 The foaming gas is nitrogen or air with traces of $C_6F_{14}$ to slow down gas diffusion between bubbles. The generated foams are then filled into the glass column of the commercial FoamScan device (TECLIS, Lyon, France) which has a square cross-section ($25\times 25~\milli\meter$) and is $200~\milli\meter$ tall. Adding surfactant solution at a constant flow rate at the top of the column (forced drainage), as sketched in Figure \ref{fig:schema}a, results in a uniform distribution of the volume fraction \cite{Weaire2001}. Its value is obtained by measuring the electrical conductivity across the column using pairs of electrodes integrated into opposing walls of the cell (Fig. \ref{fig:schema}a). This procedure rests on a general empirical relation for conductivity which is now well established and is subject to a relative experimental  error of about $5~\%$ in volume fraction $\phi$ \cite{Feitosa2005}.

  \begin{figure}[ht!]
		\includegraphics[width=8cm]{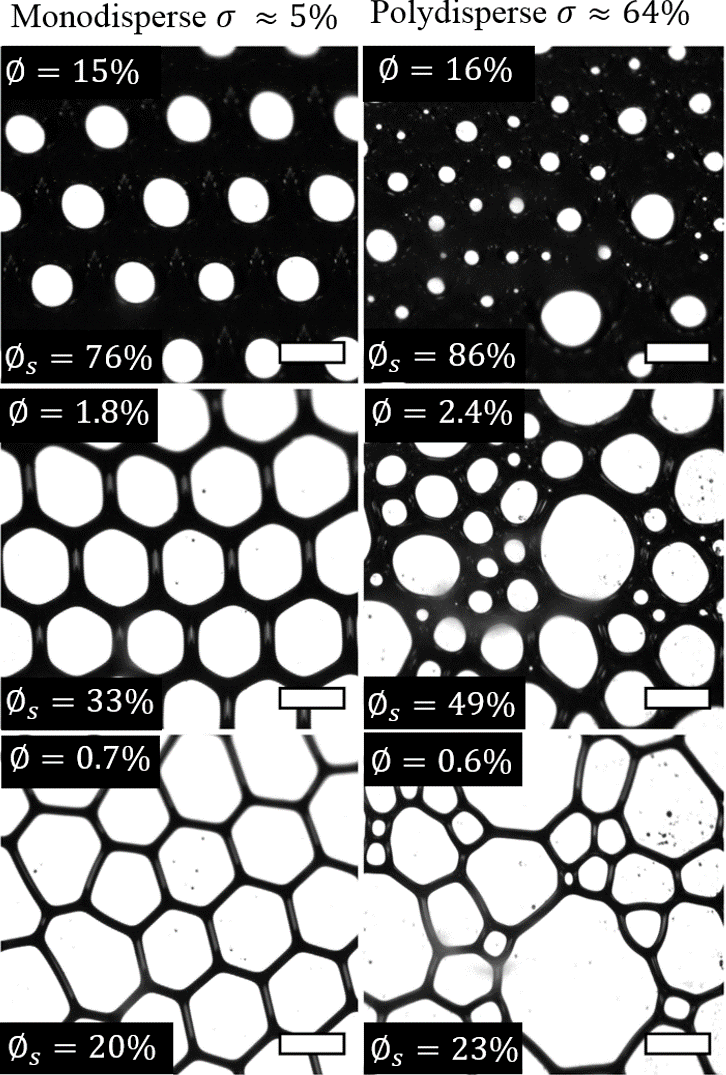} 
		\caption{Examples of photographs taken  at the surface of the foam through the prism (Fig. \ref{fig:schema}b) for three different volume fractions and for two different polydispersities $\sigma$. The scale bars are equal to $1~\milli\meter$.}
		\label{fig:photos}
\end{figure}

The surface fraction $\phi_s$ is obtained by imaging light reflected from the foam surface with a telecentric lens through a prism with a $45\degree$ angle, as shown in Figure \ref{fig:schema}b. Light reflected from the flat wetting film which separates a bubble from the glass surface is then received by the camera (and appears white), while light reflected by a curved interface is not detected by the camera (and appears black). This technique was first developed in the $1990$s  for determining bubbles size distributions \cite{Garrett1993,Mukherjee1995}. Here we expand this application to surface fraction measurements. Examples of the resulting photographs for different volume fractions and polydispersities are shown in Figure \ref{fig:photos}.  Using ImageJ software \cite{ImageJ}, these photographs are readily converted into a surface fraction $\phi_s$ by taking the ratio of the black area to the imaged surface (sec. \ref{app:surf_liq_frac} in the Supplementary Materials). The relative error made in the surface fraction measurement is of the order of $5~\%$ with a systematic tendency to underestimate $\phi_s$ due to a combination of uncertainties in the optical path selection and in the image treatment.

\section{Experimental results}
Figure \ref{fig:graph} shows the experimental data for surface fraction $\phi_s$ as a function of volume fraction $\phi$ over one order of magnitude of bubble sizes ($0.1~\milli\metre < \langle R \rangle  < 2~\milli\metre$) and a wide range of polydispersities ($1~\% < \sigma < 64~\%$). For the filled data points, measurements were taken once the foam was in equilibrium, \textit{i.e.} once the volume fraction was constant for a given flow rate $Q$ of the injected liquid (forced drainage experiments). Each data point is obtained by averaging  $\phi_s$ over 10 images, taken at a foam height of $5~\centi\meter$ from the bottom, and by averaging $\phi$ over the 10 values measured at the moment when these images were taken. The open symbols are taken under dynamic conditions, \textit{i.e.} while the volume fraction changes with time. This arises for example, once the flow rate is stopped, or upon increasing the flow rate. Here, each data point corresponds to one single measurement. The latter results are therefore more prone to errors from dynamic effects or simple lack of statistics. Nevertheless, it can be seen that the data generated by both protocols fall onto what seems to be a single curve without any systematic dependence on bubble size, polydispersity or measurement protocol. We have also added a data set obtained in 1984 by H. Princen for emulsions with much smaller droplet sizes ($\langle R \rangle \approx 5-10~\micro\metre$) and in the same range of polydispersities ($\sigma\approx 50~\%$)\cite{Princen1985}. His data set coincides with our data on foams, indicating a common underlying relationship.

\section{Modelling and interpretation}
In order to interpret this experimental observation let us consider the main pressure relations in a foam. Since we will focus on an area at fixed height we can neglect the hydrostatic pressure. Forced drainage leads to a homogeneous volume fraction throughout the foam, we thus treat the liquid pressure $P_l$ as homogeneous in the measured zone. In a disordered foam with modest polydispersity the gas pressure $P_g$   varies only slightly between bubbles \cite{Weaire2001,Cantat2013}. Since this variation is small in comparison with $P_g-P_l$, we will treat  $P_g$ as constant throughout the foam, also for the bubbles in contact with the wall. Comparison with our experimental data below seems to justify this assumption, but future work needs to quantify the consequences of this approximation. $P_g$ is also sufficiently small ($P_g < 1000$ Pa) in order to consider the gas as incompressible. This then implies a constant \textit{capillary pressure}
 $P_c=P_g-P_l$ throughout the foam.
\begin{figure}[!ht]
		\includegraphics[width=8.5cm]{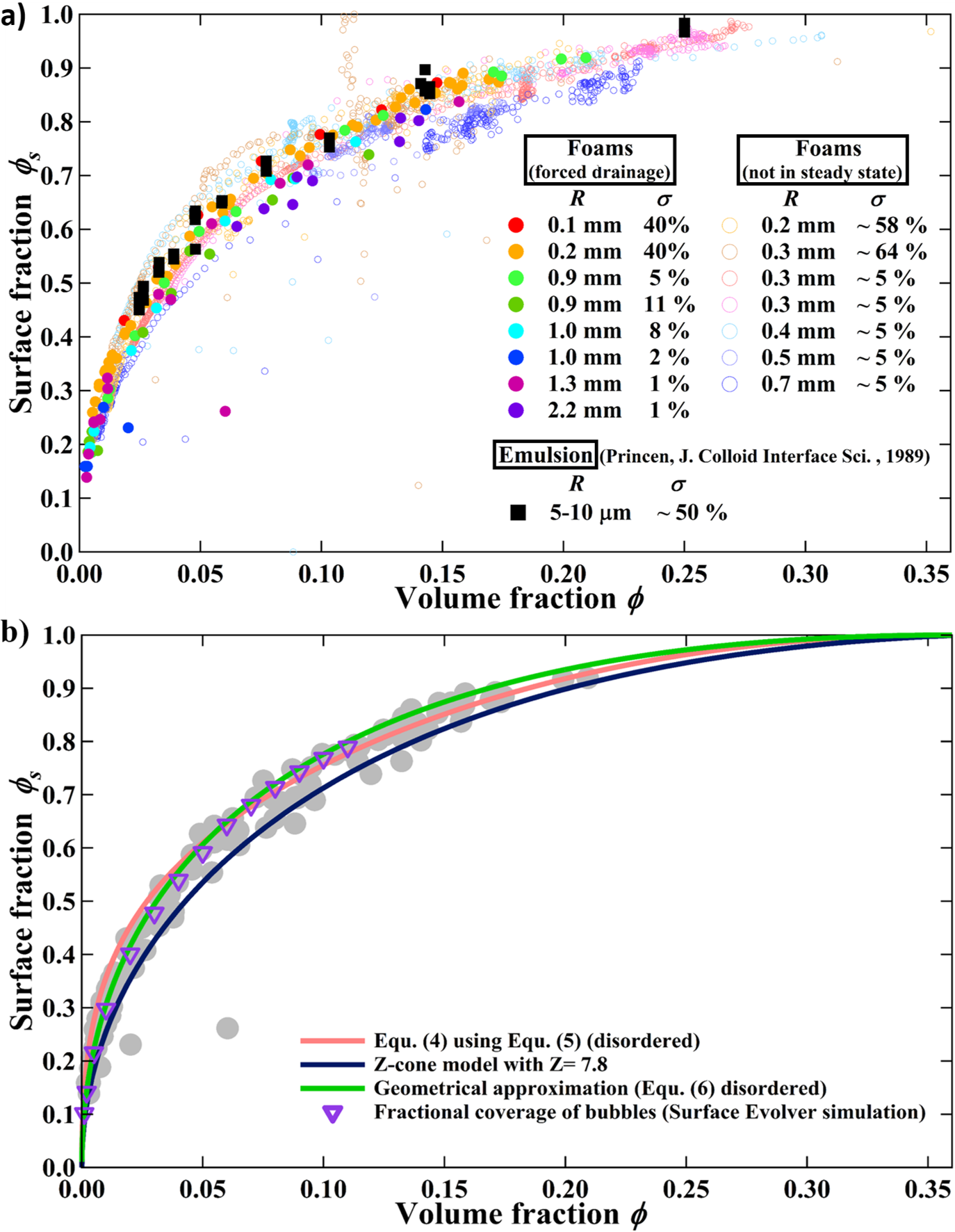} 
		\caption{(a) Experimental data relating the surface fraction $\phi_s$ to the volume fraction $\phi$ over a wide range of bubble sizes $\langle R \rangle$ ($0.1-2~\milli\metre$) and polydispersities $\sigma$ ($1-64~\%$). We have also added a data set which was obtained by Princen \cite{Princen1985} using emulsions of $\langle R \rangle \approx 5-10~\micro\metre$ and $\sigma\approx 50~\%$.  (b) A variety of theorical models are consistant with the experimental data. The relationship $\phi_s(\phi)$ also describes the fraction of the bubble surface which is wetted by the continuous phase for a kelvin foam, as obtained from Surface Evolver simulation  (data kindly provided by A. M. Kraynik and D. A. Reinelt, unpublished)}
		\label{fig:graph}
\end{figure}

The bubbles in contact with the wall create a flat film with the solid surface (which appears white in the photographs of Figure \ref{fig:photos}). This film is drained by the capillary pressure $P_c$ until $P_c$ is balanced by the repulsive force between the bubble surface and the solid wall (the so-called \textit{disjoining pressure} which arises from the presence of the surfactants \cite{Israelachvili2011}). In equilibrium, a bubble therefore exerts a force on the wall which is given by the product of the capillary pressure and the area of its contacting film. A wall in contact with a foam with a surface fraction of $\phi_s$ therefore experiences an overall pressure of $(1-\phi_s)P_c$. 

This macroscopic pressure exerted by the bubbles of a foam on a flat wall is also known as the \textit{osmotic pressure} $\Pi$, which was introduced by H. Princen \cite{Princen2001}.
  For a bulk foam it is defined as \cite{Princen2001} 
 
 \begin{equation}
\Pi =-\left[ \frac{\partial E}{\partial V} \right]_{\rm{V_{g}=const}},
\label{eq:Pi_vg_const}
\end{equation}
where $V_g$ is the gas volume, which is kept constant, and $E$ is the interfacial energy. We therefore have 

\begin{equation}
(1-\phi_s)P_c=\Pi(\phi).
\label{eq:phis_Pi}
\end{equation}
 The capillary pressure $P_c$ and osmotic pressure $\Pi$ of a foam of volume $V$ have been shown to be related \cite{Princen1988,Denkov2008,Hohler2016} via

\begin{equation}
\Pi(\phi)=(1-\phi)P_c-\frac{2}{3}\frac{E}{V}.
\label{eq:Pi_Phi}
\end{equation}

Based on our approximations we can therefore write a simple equilibrium relation for the foam bubbles which are pressed against the wall, 

\begin{equation}
\frac{(1-\phi)}{(1-\phi_s)}= 1+\frac{2}{3}\frac{E}{\Pi V}.
\label{eq:phis_phi_princen}
\end{equation}

Weaire and Hutzler stated an approximate form of this equation valid near the dry limit (Eqn. (3.42) in \cite{Weaire2001}). 
Let us note that in the case of an ordered, two-dimensional foam, Eqn. (\ref{eq:phis_phi_princen}) can be derived rigorously, leading to the same expression in which the prefactor $2/3$ is replaced by $1/2$ to account for the different dimensionality (Sec. \ref{app:2D_foam}  in the Supplementary Materials).
Equ. (\ref{eq:phis_phi_princen}) (together with Equ. (\ref{eq:Pi_vg_const})) suggests that we need to know $E(\phi)$ or $\Pi(\phi)$ for a given sample in order to relate surface to volume fraction. 
Both $E(\phi)$ and $\Pi(\phi)$ have been found to be remarkably independent of the detailed structural properties of the foam for a wide range of different samples, based on experimental and computational results (\cite{Drenckhan2015} and references therein). It follows that one may choose a representative function for either $E(\phi)$ or $\Pi(\phi)$, whereupon an estimate of the surface-volume relation can be deduced. In what follows we will highlight two possible choices.

R. H\"{o}hler and co-workers \cite{Maestro2013,Hohler2008} suggested a semi-empirical expression for the osmotic pressure

\begin{equation}
\Pi = K \frac{\gamma}{R_{32}}\frac{(\phi_c-\phi)^2}{\sqrt{\phi}},
\label{hohler_equation}
\end{equation}
where $R_{32}=\frac{\left\langle R^3 \right\rangle }{\left\langle R^2 \right\rangle }$ is the Sauter mean bubble radius. The prefactor $K$ depends on the critical volume fraction $\phi_c$ of the foam.  $K \approx 3.2$ for disordered, polydisperse foams ($\phi_c \approx 0.36$) with modest polydispersity and $K = 7.3$ for ordered, monodisperse foams ($\phi_c = 0.26$) \cite{Maestro2013,Hohler2008}. Numerical integration of Eqn. (\ref{hohler_equation}) gives the interfacial energy $E(\phi)$, and thus the evaluation of the relation between surface and volume fraction via Eqn. (\ref{eq:phis_phi_princen}). 

The resulting predictions for $\phi_s(\phi)$ for ordered and disordered foams are plotted in Figure \ref{fig:graph}b, together with a shaded representation of our data obtained from forced drainage experiments, showing that both expressions describe the data very well up to $\phi = 0.15$. For higher $\phi$, $K=3.2$ (the values corresponding to disordered foam) leads to better agreement with the data. While this is expected for the polydisperse foam samples, it may seem surprising with respect to the monodisperse foams shown in the left column of Figure \ref{fig:photos}, which seem to hint at nearly perfect order. However, this is a well-known boundary effect \cite{Meagher2015}, due to the presence of the flat wall which encourages order in its vicinity while the foam remains disordered in the bulk. In the case of polydisperse foams one might also expect that some segregation of bubble size occurs at the surface, so that its size distribution is different from that of the bulk. However, this has no effect on the derivation of Eqn. (\ref{eq:phis_phi_princen}), if the assumption of constant gas pressure remains valid.

Another expression of Eqn. (\ref{eq:phis_phi_princen}) may be obtained using the recently developed Z-cone model \cite{Hutzler2014,Whyte2015,Murtagh2015}  which approximates the geometry of a Z-faced bubble as being composed of Z equal-volume cones. This provides analytical expressions for energy and osmotic pressure as a function of volume fraction for a given number $Z$ of contacts of a bubble. These expressions are advanced and are therefore not stated explicitly here, but they can be found in the Appendix A of reference \cite{Whyte2015}. The model is derived for an ordered monodisperse foam but should provide a reasonable approximation also for the disordered polydisperse foams studied in this article with an appropriate choice of $Z$. Figure \ref{fig:graph}b shows that the variation of surface fraction with volume fraction is well described for a choice of $Z \approx 7.8$ over the entire range of data. This is a value somewhere between the average numbers of neighbours of a random monodisperse foam in the limits of low and high $\phi$, where $Z = 13.4$ and $Z = 6$, respectively. The corresponding graphs for $\phi_s(\phi)$ are shown in sec. \ref{app:Z-cone} in the Supplementary Materials.

\section{Geometrical approximation}
The comparisons in Figure \ref{fig:graph} show that there is clearly sufficiently close agreement to justify surface measurements for the estimation of the volume fraction.  
Since the use of currently available models for $E(\phi)$ and $\Pi(\phi)$ in Eqn. (\ref{eq:phis_phi_princen}) leads to rather complex expressions, we wish to close our analysis with the suggestion of a simple approximation which is based on purely geometrical arguments. These are stimulated by the fact that Equ. (\ref{eq:phis_phi_princen}) may be rewritten in terms of surface area and volume/surface fraction only, using $E=\gamma S$ and Equ. (\ref{eq:Pi_vg_const}). In the simple model of a foam as a network of liquid channels of characteristic width $w$ and characteristic length close to the bubble radius $R$ \cite{Weaire2001,Cantat2013}, the volume fraction scales as $\phi \propto w^2R/R^3 = (w/R)^2$. For the surface fraction we may write the scaling $(1-\phi_s) \propto (1-w/R)^2$. Combining both equations and requiring that $1 - \phi_s = 0$ when $\phi = \phi_c$, gives the simple relationship


\begin{equation}
1-\phi_s = \left( 1-\sqrt{\frac{\phi}{\phi_c}} \right) ^2.
\label{eq:geometric_equation}
\end{equation}

This kind of relationship has already been proposed by Princen  \cite{Princen1985,Princen2001} and Hilgenfeldt \cite{Hilgenfeldt2001}, but with slightly different pre-factors. Equ. \ref{eq:geometric_equation} is plotted in Figure \ref{fig:graph}b, showing that this provides a rather good approximation to the experimental data $-$ and to the other models. The simplicity of this relationship may seem surprising at first sight, but the fully analytical calculations (Appendix I in \cite{Princen1985}) for an ordered 2D foam gives the same relationship with power "1" instead of "2". For most practical purposes the use of this simple relationship, Eqn. \ref{eq:geometric_equation} may therefore prove sufficient.

\section{Conclusion}
In summary, our experimental measurements clearly show that there is a simple relationship between the bulk and the surface fraction of a liquid dispersion. We demonstrate that this relationship can be modelled by relating the different pressures acting in the dispersion and on its surface, leading to Eqn. \ref{eq:phis_phi_princen}. This expression is, at least roughly, independent of polydispersity and bubble size. 
A dependence on the polydispersity is expected to occur close to the critical volume fraction $\phi_c$, but the precision of our experimental set-up does not allow us to draw any quantitative conclusion. 

Taking photographs of liquid dispersions in contact with a transparent wall may therefore replace more  tedious techniques, including measurements of electrical conductivity, weight measurements or various kinds of tomography. A further advantage is also that a simple imaging technique imposes less requirements on the sample conditions (not negligeable conductivity, density contrast etc.). For most practical purposes, the simple relationship provided in Equ. (\ref{eq:geometric_equation}) may be used to relate both fractions with fairly good precision - especially in dispersions with $\phi < 0.15$ (which captures most of the commonly encountered foam samples). 

While, for simplicity, we have concentrated our experiments and the discussion on the example of foams, our arguments are valid for all liquid dispersions in which interfacial tension is constant and the relevant energy is interfacial energy. This is consistent with the agreement with data on emulsions from Princen \cite{Princen1985} shown in Figure \ref{fig:graph}a. 

A physically closely related question of great relevance to dispersion science is that of the average "surface fraction" or "wetting fraction" of the bubbles or droplets within the bulk dispersion, i.e. what fraction of the bubble/droplet surfaces is covered by the continuous phase? Or, inversely, what fraction of the bubble/droplet surfaces is not covered by the continuous phase and therefore corresponds to thin films? This is a parameter of great importance for the description of rheology and ageing phenomena in dispersions. In Figure \ref{fig:graph}b we plot this average surface fraction for a periodic Kelvin foam, obtained by Surface Evolver simulations (the data was kindly provided by A. M. Kraynik and D. A. Reinelt). We notice an excellent agreement between this data and the models derived for the surface fraction on the container wall. Moreover, expressions similar to Eqn. (\ref{eq:geometric_equation}) have already been successfully used in the modelling of bulk foam ageing \cite{Hilgenfeldt2001}. This seems to indicate that our derivation for the surface fraction on container walls may be extended, at least to a good approximation, to the derivation of the surface fraction on the bulk bubbles/droplets. Further analysis of this question is in progress.

In conclusion, in-depth investigations of dispersions in contact with solid surfaces may therefore help in answering numerous pending questions in relation to their interfacial energy (and osmotic pressure) over a wide range of volume fractions \cite{Drenckhan2015}, and, in relationship with this, the description of the non-pairwise interaction potentials between soft grains such as drops or bubbles \cite{Maestro2013,Hohler2016}.

\section{Acknowledgments}
We thank David Whyte, Reinhard H\"{o}hler, Andy Kraynik, Sascha Hilgenfeldt, Dominique Langevin and Anniina Salonen for numerous fruitful discussions. A. Salonen is thanked for M. Schneider's supervision. We thank Andy Kraynik for supplying the simulation data in Fig. \ref{fig:graph}b. We acknowledge funding from ESA (MAP AO 99-108  and contract 4000115113 Soft Matter Dynamics), CNES (through the GDR MFA) and two MPNS COST actions MP1106 "Smart and green interfaces" and MP1305 "Flowing Matter". 
WD acknowledges funding from the European Research Council (ERC Starting Grant agreement $\rm{307280}-$POMCAPS) and a public grant from the "Laboratoire d'Excellence Physics Atom Light Mater" (LabEx PALM) overseen by the French National Research Agency (ANR) as part of the "Investissements d'Avenir" program ($\rm{ANR}-\rm{10}-\rm{LABX}-\rm{0039}$, project "Soft2Hard").
 SH acknowledges that this publication has emanated from research supported in part by a research grant from  Science Foundation Ireland (SFI) under the grant number $\rm{13}/ \rm{IA}/ \rm{1926}$.



\balance


\bibliography{bibliography} 

\providecommand*{\mcitethebibliography}{\thebibliography}
\csname @ifundefined\endcsname{endmcitethebibliography}
{\let\endmcitethebibliography\endthebibliography}{}
\begin{mcitethebibliography}{29}
\providecommand*{\natexlab}[1]{#1}
\providecommand*{\mciteSetBstSublistMode}[1]{}
\providecommand*{\mciteSetBstMaxWidthForm}[2]{}
\providecommand*{\mciteBstWouldAddEndPuncttrue}
  {\def\EndOfBibitem{\unskip.}}
\providecommand*{\mciteBstWouldAddEndPunctfalse}
  {\let\EndOfBibitem\relax}
\providecommand*{\mciteSetBstMidEndSepPunct}[3]{}
\providecommand*{\mciteSetBstSublistLabelBeginEnd}[3]{}
\providecommand*{\EndOfBibitem}{}
\mciteSetBstSublistMode{f}
\mciteSetBstMaxWidthForm{subitem}
{(\emph{\alph{mcitesubitemcount}})}
\mciteSetBstSublistLabelBeginEnd{\mcitemaxwidthsubitemform\space}
{\relax}{\relax}

\bibitem[Weaire and Hutzler(2001)]{Weaire2001}
D.~L. Weaire and S.~Hutzler, \emph{{The Physics of Foams}}, Clarendon Press,
  2001\relax
\mciteBstWouldAddEndPuncttrue
\mciteSetBstMidEndSepPunct{\mcitedefaultmidpunct}
{\mcitedefaultendpunct}{\mcitedefaultseppunct}\relax
\EndOfBibitem
\bibitem[Cantat \emph{et~al.}(2013)Cantat, Cohen-Addad, Elias, Graner,
  H{\"{o}}hler, Pitois, and Rouyer]{Cantat2013}
I.~Cantat, S.~Cohen-Addad, F.~Elias, F.~Graner, R.~H{\"{o}}hler, O.~Pitois and
  F.~Rouyer, \emph{{Foams: Structure and Dynamics}}, OUP Oxford, 2013\relax
\mciteBstWouldAddEndPuncttrue
\mciteSetBstMidEndSepPunct{\mcitedefaultmidpunct}
{\mcitedefaultendpunct}{\mcitedefaultseppunct}\relax
\EndOfBibitem
\bibitem[van Hecke(2010)]{VanHecke2010}
M.~van Hecke, \emph{Journal of Physics: Condensed Matter}, 2010, \textbf{22},
  033101\relax
\mciteBstWouldAddEndPuncttrue
\mciteSetBstMidEndSepPunct{\mcitedefaultmidpunct}
{\mcitedefaultendpunct}{\mcitedefaultseppunct}\relax
\EndOfBibitem
\bibitem[H{\"{o}}hler and Cohen-Addad(2005)]{Hohler2005}
R.~H{\"{o}}hler and S.~Cohen-Addad, \emph{Journal of Physics: Condensed
  Matter}, 2005, \textbf{17}, R1041--R1069\relax
\mciteBstWouldAddEndPuncttrue
\mciteSetBstMidEndSepPunct{\mcitedefaultmidpunct}
{\mcitedefaultendpunct}{\mcitedefaultseppunct}\relax
\EndOfBibitem
\bibitem[Princen(CRC Press, 2001)]{Princen2001}
H.~M. Princen, \emph{{The Structure, Mechanics, and Rheology of Concentrated
  Emulsions and Fluid Foams}}, CRC Press, 2001\relax
\mciteBstWouldAddEndPuncttrue
\mciteSetBstMidEndSepPunct{\mcitedefaultmidpunct}
{\mcitedefaultendpunct}{\mcitedefaultseppunct}\relax
\EndOfBibitem
\bibitem[Bink(1998)]{Bink1998}
B.~Bink, \emph{{Modern Aspects of Emulsion Science}}, 1998, vol. 1998, p.
  430\relax
\mciteBstWouldAddEndPuncttrue
\mciteSetBstMidEndSepPunct{\mcitedefaultmidpunct}
{\mcitedefaultendpunct}{\mcitedefaultseppunct}\relax
\EndOfBibitem
\bibitem[Feitosa \emph{et~al.}(2005)Feitosa, Marze, Saint-Jalmes, and
  Durian]{Feitosa2005}
K.~Feitosa, S.~Marze, A.~Saint-Jalmes and D.~J. Durian, \emph{Journal of
  Physics: Condensed Matter}, 2005, \textbf{17}, 6301--6305\relax
\mciteBstWouldAddEndPuncttrue
\mciteSetBstMidEndSepPunct{\mcitedefaultmidpunct}
{\mcitedefaultendpunct}{\mcitedefaultseppunct}\relax
\EndOfBibitem
\bibitem[Hutzler \emph{et~al.}(1995)Hutzler, Verbist, Weaire, and
  Steen]{Hutzler1995}
S.~Hutzler, G.~Verbist, D.~Weaire and J.~a. V.~D. Steen, \emph{Europhysics
  Letters (EPL)}, 1995, \textbf{31}, 497--502\relax
\mciteBstWouldAddEndPuncttrue
\mciteSetBstMidEndSepPunct{\mcitedefaultmidpunct}
{\mcitedefaultendpunct}{\mcitedefaultseppunct}\relax
\EndOfBibitem
\bibitem[Vera \emph{et~al.}(2001)Vera, Saint-Jalmes, and Durian]{Vera2001}
M.~U. Vera, a.~Saint-Jalmes and D.~J. Durian, \emph{Applied optics}, 2001,
  \textbf{40}, 4210--4214\relax
\mciteBstWouldAddEndPuncttrue
\mciteSetBstMidEndSepPunct{\mcitedefaultmidpunct}
{\mcitedefaultendpunct}{\mcitedefaultseppunct}\relax
\EndOfBibitem
\bibitem[Sol{\'{o}}rzano \emph{et~al.}(2013)Sol{\'{o}}rzano, Pardo-Alonso,
  de~Saja, and Rodr{\'{\i}}guez-P{\'{e}}rez]{Solorzano2013}
E.~Sol{\'{o}}rzano, S.~Pardo-Alonso, J.~de~Saja and
  M.~Rodr{\'{\i}}guez-P{\'{e}}rez, \emph{Colloids and Surfaces A:
  Physicochemical and Engineering Aspects}, 2013, \textbf{438}, 159--166\relax
\mciteBstWouldAddEndPuncttrue
\mciteSetBstMidEndSepPunct{\mcitedefaultmidpunct}
{\mcitedefaultendpunct}{\mcitedefaultseppunct}\relax
\EndOfBibitem
\bibitem[Drenckhan and Hutzler(2015)]{Drenckhan2015}
W.~Drenckhan and S.~Hutzler, \emph{Advances in Colloid and Interface Science},
  2015, \textbf{224}, 1--16\relax
\mciteBstWouldAddEndPuncttrue
\mciteSetBstMidEndSepPunct{\mcitedefaultmidpunct}
{\mcitedefaultendpunct}{\mcitedefaultseppunct}\relax
\EndOfBibitem
\bibitem[Cheng and Lemlich(1983)]{Cheng1983}
H.~C. Cheng and R.~Lemlich, \emph{Industrial {\&} Engineering Chemistry
  Fundamentals}, 1983, \textbf{22}, 105--109\relax
\mciteBstWouldAddEndPuncttrue
\mciteSetBstMidEndSepPunct{\mcitedefaultmidpunct}
{\mcitedefaultendpunct}{\mcitedefaultseppunct}\relax
\EndOfBibitem
\bibitem[Wang and Neethling(2009)]{Wang2009}
Y.~Wang and S.~J. Neethling, \emph{Colloids and Surfaces A: Physicochemical and
  Engineering Aspects}, 2009, \textbf{339}, 73--81\relax
\mciteBstWouldAddEndPuncttrue
\mciteSetBstMidEndSepPunct{\mcitedefaultmidpunct}
{\mcitedefaultendpunct}{\mcitedefaultseppunct}\relax
\EndOfBibitem
\bibitem[Garrett \emph{et~al.}(1993)Garrett, Hines, Joyce, and
  Whittal]{Garrett1993}
P.~Garrett, J.~Hines, S.~Joyce and P.~Whittal, \emph{Report prepared for
  Univlever R \& D}, 1993\relax
\mciteBstWouldAddEndPuncttrue
\mciteSetBstMidEndSepPunct{\mcitedefaultmidpunct}
{\mcitedefaultendpunct}{\mcitedefaultseppunct}\relax
\EndOfBibitem
\bibitem[Mukherjee and Wiedersich(1995)]{Mukherjee1995}
S.~Mukherjee and H.~Wiedersich, \emph{Colloids and Surfaces A: Physicochemical
  and Engineering Aspects}, 1995, \textbf{95}, 159--172\relax
\mciteBstWouldAddEndPuncttrue
\mciteSetBstMidEndSepPunct{\mcitedefaultmidpunct}
{\mcitedefaultendpunct}{\mcitedefaultseppunct}\relax
\EndOfBibitem
\bibitem[Ima()]{ImageJ}
\url{https://imagej.nih.gov/ij/}, ImageJ website\relax
\mciteBstWouldAddEndPuncttrue
\mciteSetBstMidEndSepPunct{\mcitedefaultmidpunct}
{\mcitedefaultendpunct}{\mcitedefaultseppunct}\relax
\EndOfBibitem
\bibitem[Princen(1985)]{Princen1985}
H.~M. Princen, \emph{Journal of Colloid and Interface Science}, 1985,
  \textbf{105}, 150--171\relax
\mciteBstWouldAddEndPuncttrue
\mciteSetBstMidEndSepPunct{\mcitedefaultmidpunct}
{\mcitedefaultendpunct}{\mcitedefaultseppunct}\relax
\EndOfBibitem
\bibitem[Israelachvili(2011)]{Israelachvili2011}
J.~N. Israelachvili, \emph{{Intermolecular and Surface Forces: Revised Third
  Edition}}, 2011, p. 704\relax
\mciteBstWouldAddEndPuncttrue
\mciteSetBstMidEndSepPunct{\mcitedefaultmidpunct}
{\mcitedefaultendpunct}{\mcitedefaultseppunct}\relax
\EndOfBibitem
\bibitem[Princen(1988)]{Princen1988}
H.~M. Princen, \emph{Langmuir}, 1988, \textbf{4}, 164--169\relax
\mciteBstWouldAddEndPuncttrue
\mciteSetBstMidEndSepPunct{\mcitedefaultmidpunct}
{\mcitedefaultendpunct}{\mcitedefaultseppunct}\relax
\EndOfBibitem
\bibitem[Denkov \emph{et~al.}(2008)Denkov, Tcholakova, Golemanov, Hu, and
  Lips]{Denkov2008}
N.~D. Denkov, S.~Tcholakova, K.~Golemanov, T.~Hu and A.~Lips, \emph{AIP
  Conference Proceedings}, 2008, \textbf{1027}, 902--904\relax
\mciteBstWouldAddEndPuncttrue
\mciteSetBstMidEndSepPunct{\mcitedefaultmidpunct}
{\mcitedefaultendpunct}{\mcitedefaultseppunct}\relax
\EndOfBibitem
\bibitem[H{\"{o}}hler()]{Hohler2016}
R.~H{\"{o}}hler, \emph{private communication}\relax
\mciteBstWouldAddEndPuncttrue
\mciteSetBstMidEndSepPunct{\mcitedefaultmidpunct}
{\mcitedefaultendpunct}{\mcitedefaultseppunct}\relax
\EndOfBibitem
\bibitem[Maestro \emph{et~al.}(2013)Maestro, Drenckhan, Rio, and
  H{\"{o}}hler]{Maestro2013}
A.~Maestro, W.~Drenckhan, E.~Rio and R.~H{\"{o}}hler, \emph{Soft Matter}, 2013,
  \textbf{9}, 2531\relax
\mciteBstWouldAddEndPuncttrue
\mciteSetBstMidEndSepPunct{\mcitedefaultmidpunct}
{\mcitedefaultendpunct}{\mcitedefaultseppunct}\relax
\EndOfBibitem
\bibitem[H{\"{o}}hler \emph{et~al.}(2008)H{\"{o}}hler, Sang, Lorenceau, and
  Cohen-Addad]{Hohler2008}
R.~H{\"{o}}hler, Y.~Y.~C. Sang, E.~Lorenceau and S.~Cohen-Addad,
  \emph{Langmuir}, 2008, \textbf{24}, 418--425\relax
\mciteBstWouldAddEndPuncttrue
\mciteSetBstMidEndSepPunct{\mcitedefaultmidpunct}
{\mcitedefaultendpunct}{\mcitedefaultseppunct}\relax
\EndOfBibitem
\bibitem[Meagher \emph{et~al.}(2015)Meagher, Whyte, Banhart, Hutzler, Weaire,
  and Garc{\'{\i}}a-Moreno]{Meagher2015}
A.~J. Meagher, D.~Whyte, J.~Banhart, S.~Hutzler, D.~Weaire and
  F.~Garc{\'{\i}}a-Moreno, \emph{Soft Matter}, 2015, \textbf{11}, 4710--6\relax
\mciteBstWouldAddEndPuncttrue
\mciteSetBstMidEndSepPunct{\mcitedefaultmidpunct}
{\mcitedefaultendpunct}{\mcitedefaultseppunct}\relax
\EndOfBibitem
\bibitem[Hutzler \emph{et~al.}(2014)Hutzler, Murtagh, Whyte, Tobin, and
  Weaire]{Hutzler2014}
S.~Hutzler, R.~P. Murtagh, D.~Whyte, S.~T. Tobin and D.~Weaire, \emph{Soft
  Matter}, 2014, \textbf{10}, 7103--8\relax
\mciteBstWouldAddEndPuncttrue
\mciteSetBstMidEndSepPunct{\mcitedefaultmidpunct}
{\mcitedefaultendpunct}{\mcitedefaultseppunct}\relax
\EndOfBibitem
\bibitem[Whyte \emph{et~al.}(2015)Whyte, Murtagh, Weaire, and
  Hutzler]{Whyte2015}
D.~Whyte, R.~Murtagh, D.~Weaire and S.~Hutzler, \emph{Colloids and Surfaces A:
  Physicochemical and Engineering Aspects}, 2015, \textbf{473}, 115--122\relax
\mciteBstWouldAddEndPuncttrue
\mciteSetBstMidEndSepPunct{\mcitedefaultmidpunct}
{\mcitedefaultendpunct}{\mcitedefaultseppunct}\relax
\EndOfBibitem
\bibitem[Murtagh \emph{et~al.}(2015)Murtagh, Whyte, Weaire, and
  Hutzler]{Murtagh2015}
R.~Murtagh, D.~Whyte, D.~Weaire and S.~Hutzler, \emph{Philosophical Magazine},
  2015, \textbf{95}, 4023--4034\relax
\mciteBstWouldAddEndPuncttrue
\mciteSetBstMidEndSepPunct{\mcitedefaultmidpunct}
{\mcitedefaultendpunct}{\mcitedefaultseppunct}\relax
\EndOfBibitem
\bibitem[Hilgenfeldt \emph{et~al.}(2001)Hilgenfeldt, Koehler, and
  Stone]{Hilgenfeldt2001}
S.~Hilgenfeldt, S.~Koehler and H.~Stone, \emph{Physical review letters}, 2001,
  \textbf{86}, 4704--4707\relax
\mciteBstWouldAddEndPuncttrue
\mciteSetBstMidEndSepPunct{\mcitedefaultmidpunct}
{\mcitedefaultendpunct}{\mcitedefaultseppunct}\relax
\EndOfBibitem
\bibitem[Gaillard \emph{et~al.}(2015)Gaillard, Honorez, Jumeau, Elias, and
  Drenckhan]{Gaillard2015}
T.~Gaillard, C.~Honorez, M.~Jumeau, F.~Elias and W.~Drenckhan, \emph{Colloids
  and Surfaces A: Physicochemical and Engineering Aspects}, 2015, \textbf{473},
  68--74\relax
\mciteBstWouldAddEndPuncttrue
\mciteSetBstMidEndSepPunct{\mcitedefaultmidpunct}
{\mcitedefaultendpunct}{\mcitedefaultseppunct}\relax
\EndOfBibitem
\end{mcitethebibliography}
\bibliographystyle{rsc} 

\twocolumn[
  \begin{@twocolumnfalse}
\vspace{5cm}
\sffamily
\begin{tabular}{m{6cm} p{13.5cm} }

\vspace{0.3cm} & \noindent\LARGE{\textbf{Supplementary Material}} \\
\vspace{3cm} & \vspace{3cm} \\



\end{tabular}
 \end{@twocolumnfalse} \vspace{0.6cm}
]








\appendix
\setcounter{figure}{0}
\section{\label{app:generation}  Foam generation} 
We use multiple technique to generate our foams. For polydisperse foam, we use three methods. The FoamScan which is a commercial device (TECLIS, Lyon, France) is a $200~\milli\meter$ tall column with a square cross-section  of $25 \times 25~\milli\meter$. At the bottom, there is a  porous disc which can be used to generate the foam. The gas is injected into the porous disc  which is submerged in the solution. Bubbles rise from the bottom to create the foam. With this method we create foams with a polydispersity $\sigma \approx~5-10~\%$ and a bubble size $<R>~\approx~1~\milli\meter$. 
The foam can also be created by shaking the liquid in a closed container. This technique gives very polydisperse foam ($\langle R\rangle \approx 3~\milli\meter$, $ \sigma \approx 60~\%$). We also use the double syringe method \cite{Drenckhan2015} in which the gas and the solution are placed in a syringe linked to a second one. The foam is generated by pushing back and forth. this technique gives a bubbles size $<R>\approx 0.1 \milli\meter$ and a polydispersity $\sigma \approx 40\%$.

For monodisperse foams, we use two techniques. The simplest one is to inject the gas in the solution with a needle (with a diameter of $0.4~\milli\meter$) which is placed at the bottom of the FoamScan column. The bubble radius is then of $0.3~\milli\meter$ and the polydispersity is around $5~\%$.
The second method is a microfluidic technique which is the flow focusing though a T-junction. The solution is injected in the first branch and the gas in the second one. They meet at the junction where they create equal-volume bubbles which flow out of the third branch. The FoamScan is filled with this foam through an opening at the top of the column. With this method, it is possible to generate a wide range of bubble sizes ($1-2~\milli\meter$)  with a polydispersity of $1-2~\%$. 

\section{\label{app:size_poly} Bubble size and polydispersity}  

\begin{figure}[!ht]
	\center
	\includegraphics[width=0.45\textwidth]{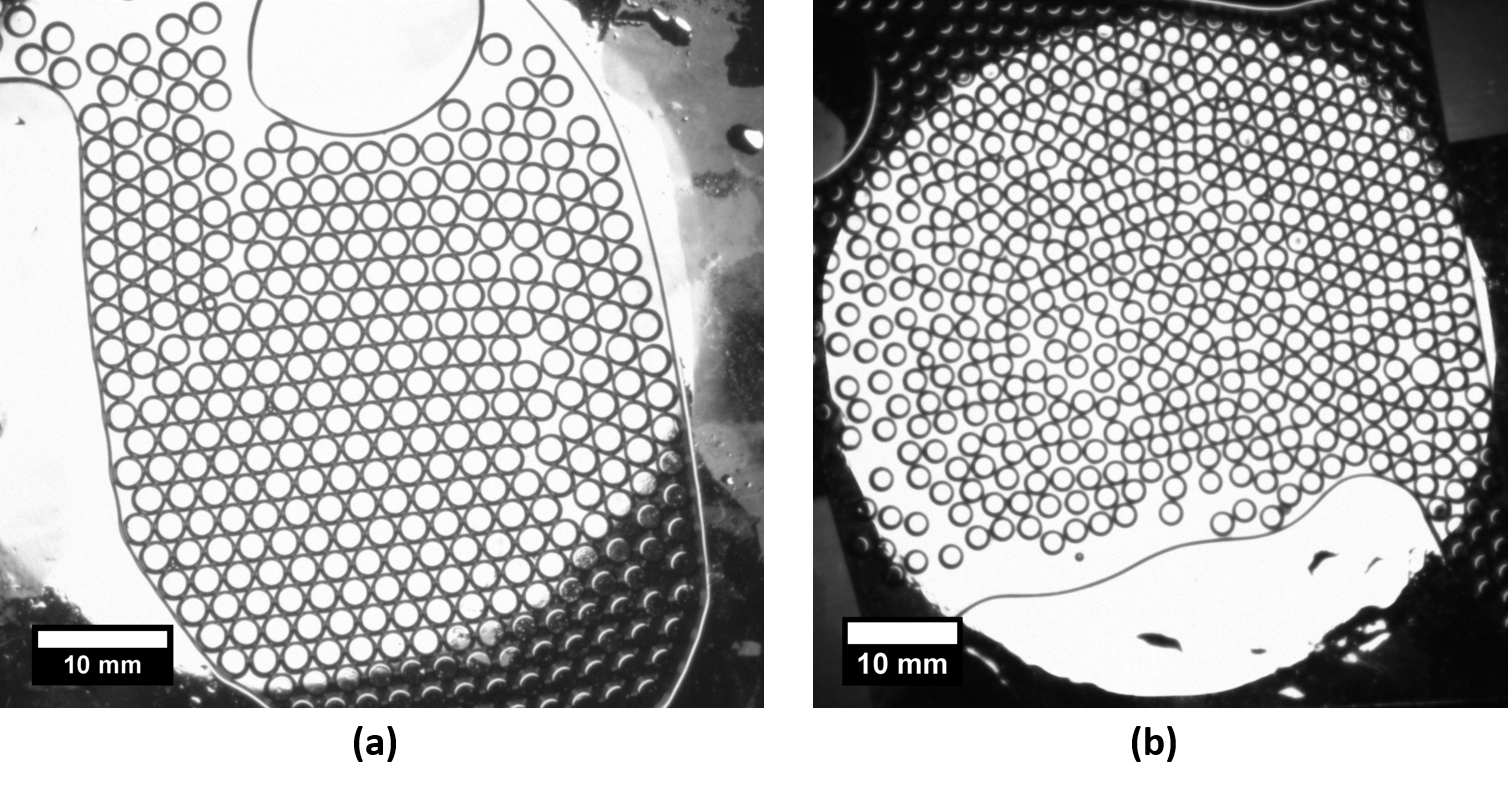} 
	\caption{Photographs of bubbles between two glass plates taken with a camera placed on top of the sample and enlighten with a circular light source below. (a) bubbles size of $\langle R \rangle=2.2~\milli\metre$ and a polydispersity of $1~\%$. (b) Bubbles size of $\langle R \rangle =1~\milli\metre$ and a polydispersity of $8~\%$.}
	\label{fig:bubble_size}	
\end{figure}

To measure the bubble size and the polydispersity of the foam,  we take a sample in the column of the FoamScan and dilute it with the same solution between two glass plates which are separated by a known gap which is smaller than the bubble diameter. Diluting the foam sample in   allows to separate the bubbles from each other and to see all of them individually as shown in Fig. \ref{fig:bubble_size}.
A camera is positioned above the sample which is illuminated from below with a circular light source to take pictures of the bubbles. 
On these pictures, the edges of the bubbles appear in black (see Fig. \ref{fig:bubble_size}).
With a basic image processing in ImageJ software \cite{ImageJ}, we obtain the inner radius of the bubbles, $R_{c}$, as described in detail in \cite{Gaillard2015}. Knowing the distance between the light and the sample, $L_l$, the radius of the light, $R_l$, the gap, $h$, between the two glass plates, the optical index, $n$, and the inner radius of the bubbles, we can calculate the real radius of the bubbles with  Equ. (17) of \cite{Gaillard2015}.

Based on these pictures and on the calculation of the radii, we measure the polydispersity of the foam  given by 
$\sigma= \frac{\sqrt{\left\langle R^2\right\rangle -\left\langle R\right\rangle ^2}}{\left\langle R \right\rangle}$.
Hence, for the figure \ref{fig:bubble_size}a, $\langle R \rangle =2.2~\milli\metre$ and $\sigma=1~\%$ while for the  picture  \ref{fig:bubble_size}b, $\langle R \rangle =1~\milli\metre$ and $\sigma=8~\%$.
\section{\label{app:surf_liq_frac}  Measurement of the surface liquid fraction} 
The column of the FoamScan is provided with prisms which are glued on the container surface between successive electrodes as sketched in Fig. \ref{fig:schema}a.  Pictures of the bubbles at the wall are taken though the prism (Fig. \ref{fig:schema}b) using a camera with a telecentric lens.
 
 
To obtain a sharp image of the bubbles at the wall, the angles between the container surface and the light and between the container surface and the camera must be positioned at an angle of $45~\degree$. That way the maximum intensity will arrive in the prism. Additionaly, we use a telecentric lens so that  only the rays 
reaching the prism with an angle of $45\degree$ will be seen by the camera (Fig. \ref{fig:schema}b in the main article). 
\begin{figure}[!ht]
	\center
	\begin{subfigure}[b]{0.23\textwidth}
	\includegraphics[width=\textwidth]{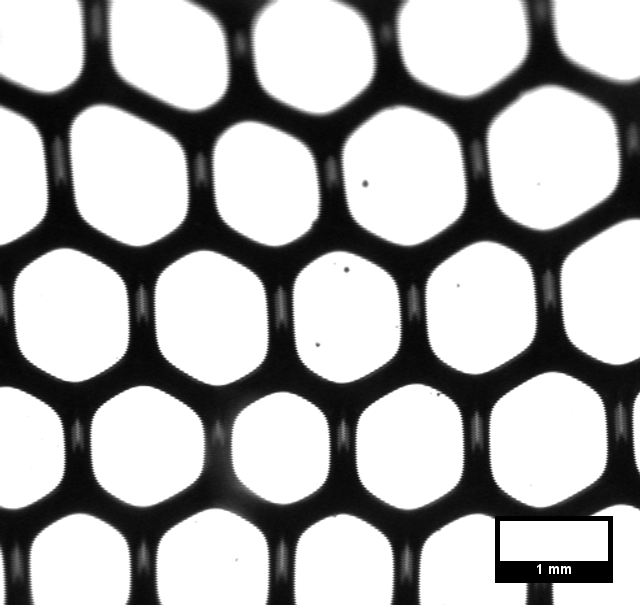}
	\caption{}
	\label{fig:original}
	\end{subfigure}
	\begin{subfigure}[b]{0.23\textwidth} 
	\includegraphics[width=\textwidth]{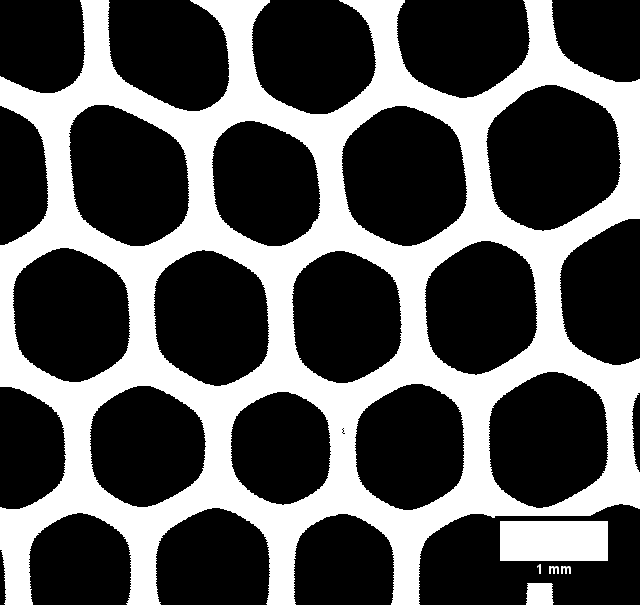} 
	\caption{}
	\label{fig:binaire}
	\end{subfigure}
	\caption{Photographs of bubbles of the surface at the container wall. (a) Original photograph taken with a camera and a telecentric lens as shown in Fig. \ref{fig:schema}b. (b) Same photograph which has been binarised and inverted.}
	\label{fig:surface}
\end{figure}
So, if the light beam arrives on a plane surface (film against the wall), the light will be reflected in the axis of the camera and the film  will appear in white on the photograph (Fig. \ref{fig:original}). On the contrary, if the beams is reflected by a curved  surface,  it will be reflected with another angle and will not be seen by the camera. This point will  thus appear in black in the photograph (Fig. \ref{fig:original}). We therefore obtain a picture which is black, where the liquid wets the surface and white where the bubbles are pressed against the wall.
We use the free software ImageJ to binarise and to invert the photographs (Fig. \ref{fig:binaire}). The surface fraction is then calculated with the ratio of the number of black pixels over the total number of pixels. 
Fig. \ref{fig:surface} shows an example of the surface of a foam with a volume fraction of $\phi =1.8~\%$. For this foam the surface fraction is $\phi_s=34~\%$.
This measurement technique has the systematic tendency to underestimate slightly the surface fraction. On the one hand, the telecentric lens allows for a small variation in the angle of the detected light rays. On the other hand, the threshold, which is used to turn the photograph into a black-and-white picture has the tendency to cut slightly the boundary of the black zones. We estimate that the combination of both errors leads to an underestimation up to $5~\%$ of the surface liquid fraction.

\section{ \label{app:2D_foam} Calculation of Equ. (\ref{eq:geometric_equation}) for a 2D foam}
In the case of a two-dimensional hexagonal foam one can derive rigourously the foam energy \cite{Princen2001}

\begin{figure}[!ht]
    \center
    \includegraphics[width=0.49\textwidth]{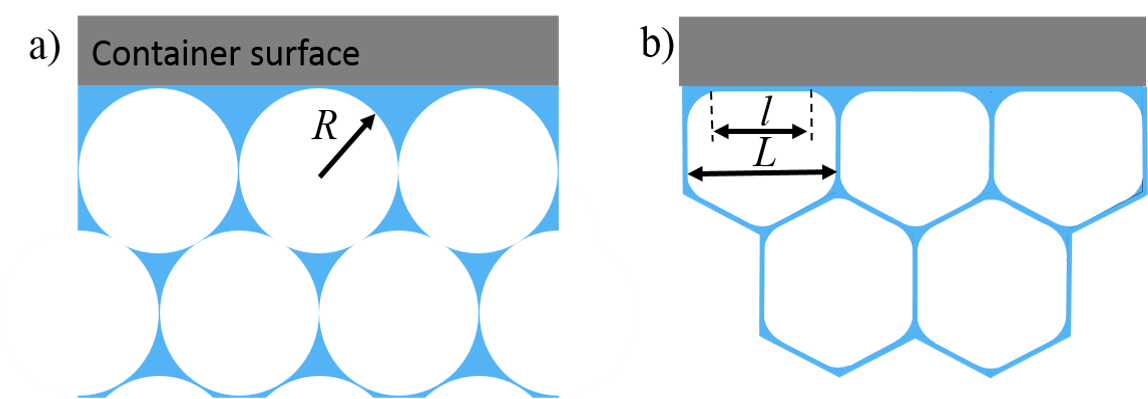}
    \caption{2D foam (a) with undeformed bubbles of radius $R$ and (b) with hexagonal shape bubbles of width $L$ and length of film against the wall $l$.}
    \label{fig:Calcul_2D}
\end{figure}

\begin{equation}
E(\phi)= \gamma S(\phi) = \gamma S_0 \frac{( 1-\phi^{1/2}\phi_c^{1/2}) }{(1-\phi_c)^{1/2}(1-\phi)^{1/2}},
\label{eq:princen_2000}
\end{equation}
where $S_0 = 2 \pi R$ is the surface of the undeformed bubbles, which corresponds to a line length in two dimensions. $R$ is the radius of the undeformed bubbles (Fig. \ref{fig:Calcul_2D}a), $\gamma$ the line tension, and $\phi$ the area liquid fraction. Similarly, one obtains for the two-dimensional osmotic pressure \cite{Princen2001}

\begin{equation}
\Pi(\phi)=\frac{\gamma}{R}\frac{(1-\phi)^{1/2}}{(1-\phi_c)^{1/2}}[ \left( \frac{\phi_c}{\phi}\right) ^{1/2}-1 ].
\label{eq:osmotic_pressure_2000}
\end{equation}
The surface fraction in a 2D-foam (Fig. \ref{fig:Calcul_2D}b) is given by (Appendix I in \cite{Princen1985})

\begin{equation}
\frac{l}{L}=(1-\phi_s)=1-\sqrt{\frac{\phi}{\phi_c}},
\label{eq:surf_frac_liq_2D}
\end{equation}
with $l$ being the length of film against the wall and $L$ the width of the bubble against the wall (Fig. \ref{fig:Calcul_2D}b).
Combining Equ.s (\ref{eq:princen_2000})- (\ref{eq:surf_frac_liq_2D}) gives
\begin{equation}
1+\frac{1}{2}\frac{E}{\Pi A}=\frac{1-\phi}{1-\phi_s}
\label{eq:surf_liq_frac_bulk_liq_frac_2D}
\end{equation}
which is the 2D analogy of Equ. (\ref{eq:phis_phi_princen}) of the main article with  $2/3$ replaced by $1/2$. Here $A$ is the area of the foam, which is given by $A = A_B (1 - \phi)$ with $A_B$ being the bubble area. 

\section{\label{app:Z-cone}  Comparison with the Z-cone model}  


The Z-cone model can be used to calculate the value of the osmotic pressure and the energy of the foam which are necessary to evaluate the Equ. (\ref{eq:phis_phi_princen}). In this model a key parameter is $Z$  which is the number of faces of a bubble. 

 \begin{figure}[!ht]
	\center
	 \includegraphics[width=0.48\textwidth]{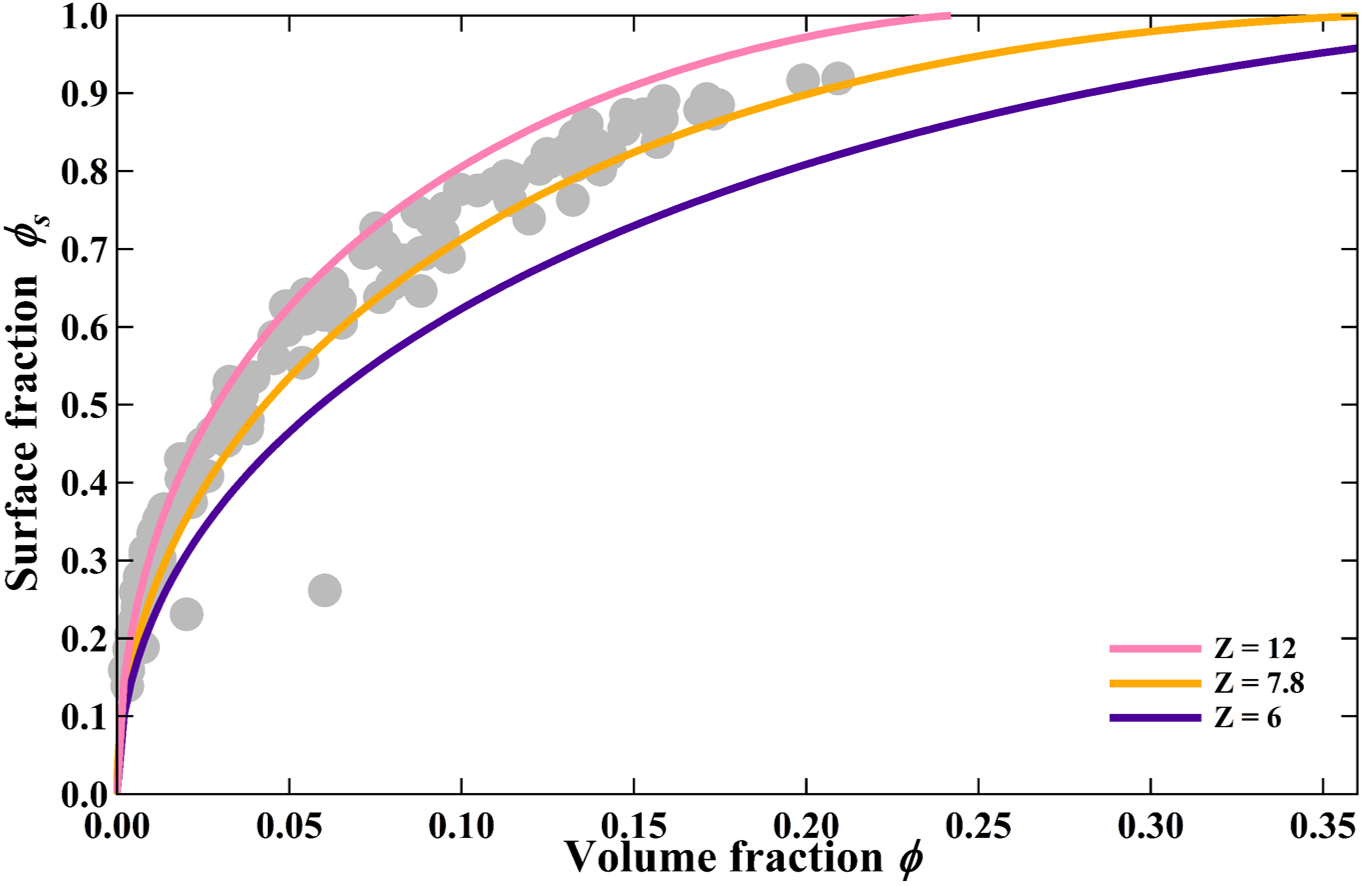} 
	\caption{Surface fraction versus volume fraction for the Z-cone model with 3 values of $Z$ ($Z=6, 7.8$ and $12$). The gray points are the experimental data extracted from Figure \ref{fig:graph}a in the main article.}
	\label{fig:Z-cone_model}
\end{figure}
$Z$ is derived for an ordered monodisperse foam but the model can be used for disordered polydisperse foams. Since $Z$ represents the number of equal-sized cones or faces of the bubbles in the foam, it can also be seen as the number of neighbours of the bubble. 
For example, if $Z=6$ it means that the bubbles have 6 neighbours and for bubbles with 12 neighbours $Z= 12$.

Fig. \ref{fig:Z-cone_model} shows  the surface fraction, $\phi_s$, versus the volume fraction, $\phi$, $Z=6$ in purple,  $Z=12$ in pink and $Z=7.8$ in orange which is the values of $Z$ allowing the better agreement with the data. The gray dots are the experimental data points (static foam in fig. \ref{fig:graph}a). For $Z=12$, there is a good agreement between the  Z-cone model and the experimental points for small bulk liquid fractions. It is due to the fact that a bubble has a lot of neighbours in a dry foam. 
It is noticeable  that a unique value of $Z$ ($Z=7.8$) allows to describe the entire range of data. This value is between the average neighbours of a random monodisperse foam in the limits of low and high bulk liquid fraction.



\balance


\end{document}